\documentclass[12pt]{article}
\usepackage{euscript,amssymb,amsmath,graphicx}

\DeclareMathOperator{\Bd}{Bd}
\newcommand{\rnum}[1]{\mathrm{I\!R^#1}}
\newcommand{\kartiv}[3]{\begin{figure}[t]\begin{center}
\includegraphics[width=#1]{#2}
\end{center}\caption{#3}
\end{figure}}
\newcommand{\ssy}[5]{#1, \emph{#2} \textbf{#3}, #5 (#4).}
\newcommand{\rmd}{\mathrm{d}}
\newcommand{\dalamb}{\raise-0.1em \hbox{$\Box$}}
\newcounter{par}
\newcommand{\parnum}[1]{\par\smallskip\noindent\refstepcounter{par}%
\textbf{\thepar. #1}}
\title{The wormhole hazard}
\author{S. Krasnikov\thanks{Email: redish@gao.spb.ru}}
\begin{document}
\maketitle
\begin{abstract}
To predict the outcome of (almost) any experiment we have to
assume that our spacetime is globally hyperbolic. The wormholes, if
they exist, cast doubt on the validity of this assumption. At the
same time, no evidence has been found so far (either
observational, or theoretical) that the possibility of their existence
can be safely neglected.
\end{abstract}
\section{Introduction}
According to a widespread belief general relativity is the science of
gravitational force. Which means, in fact, that it is important only
in cosmology, or in extremely subtle effects involving tiny post-Newtonian
corrections.

However, this point of view is, perhaps, somewhat simplistic. Being
concerned with the structure of spacetime itself, relativity sometimes
poses problems vital to
whole physics. Two best known examples are singularities and time
machines. In this talk I discuss another, a little less known, but, in
my belief, equally important problem (closely related to the preceding
two). In a nutshell it can be formulated in the form of two question:
What principle must be added to general relativity to provide it (and
all other physics along with it) by predictive power? Does not
the (hypothetical) existence of wormholes endanger that (hypothetical)
principle?

\section{Global hyperbolicity and predictive power}
\subsection{Globally hyperbolic spacetimes}
The globally hyperbolic spacetimes are the spacetimes with especially
simple and benign causal structure, the spacetimes where we can use
physical theories in the same manner as is customary in the Minkowski
space.
\parnum{Definition.}\label{def} A spacetime $M$ is  \emph{globally
hyperbolic} if it contains a subset $\mathcal S$ (called a \emph{Cauchy
surface}) which is met exactly once by every inextendible causal curve
in $M$.
\parnum{Remark}\label{rem:intr}
We shall not need the concept of a globally hyperbolic \emph{subset}
of a spacetime, so below, whenever I call a subset $U$ of $M$ globally
hyperbolic\footnote{A more rigorous term would be \emph{intrinsically
globally hyperbolic} \cite{eotm}.}, I mean that $U$ is globally
hyperbolic, when considered as an (extendible) spacetime by itself,
not as a part of $M$.

\noindent Topologically globally hyperbolic spacetimes
are `dull'~\cite{HawEl} in the following sense.
\parnum{Property.}\label{prop1}
All  Cauchy
surfaces of a given globally
hyperbolic spacetime $M$ are homeomorphic. Moreover,
$$
M=\mathcal  S\times\rnum{1},
$$
where $\mathcal  S$ is a
three-dimensional manifold, and $\mathcal  S\times\{x\}$ are  Cauchy
surfaces of $M$ for all $x\in\rnum{1}$.

\noindent The causal structure of globally hyperbolic spacetimes is also nice.
\parnum{Property.}\label{prop2} Globally hyperbolic  spacetimes
do not contain closed causal curves  (for, if such a curve $\ell$
meets a set $\mathcal P$ once, then the  causal curve
$\ell\circ\ell$ meets $\mathcal P$ twice).

\parnum{Examples.} The Minkowski and  Friedmann spacetimes
are both globally hyperbolic. And a `conical' spacetime
$$
\rmd s^2= -\rmd t^2 + \rmd x^2 + \rmd r^2 + r^2 \rmd \varphi^2,
\qquad r>0,\ \varphi\in[0,\varphi_0<2\pi),
$$
which describes a cosmic string, is not. For, consider a ball $B=
\{p\colon\: r(p)\leqslant\pi/2\}$. By definition a  Cauchy surface must be
\emph{achronal} (i.~e.\ no its points can be connected by a timelike
curve).  But the $t$-coordinates
of all points of $\mathcal P\cap B$, where $\mathcal P$ is achronal
and passes through the origin of the coordinate system, are
bounded from above by some $t_0>0$. So, the inextendible causal
curve
$$
x=\phi=0,\quad t= 2t_0+\tan  r
$$
will never intersect $\mathcal P$. Whence $\mathcal
P$ is not a Cauchy surface.

Note that global hyperbolicity is not directly related to the absence
of singularities. For example, the (two-dimensional) anti-de~Sitter
spacetime, which is a strip
$$
\rmd s^2=\sec^2x\,(-\rmd t^2 + \rmd x^2),\qquad -\pi/2<x<\pi/2
$$
(see Fig.~\ref{fig:3}a), is non-globally hyperbolic, though it is
singularity free (the lines $x=\pm\pi/2$ are infinities rather than
singularities --- it would take infinitely large time for an observer
moving with a finite acceleration to reach them). At the same time the
Schwarzschild spacetime (see Fig.~\ref{fig:3}b) \emph{is} globally
hyperbolic in spite of the singularities.

\kartiv{30pc}{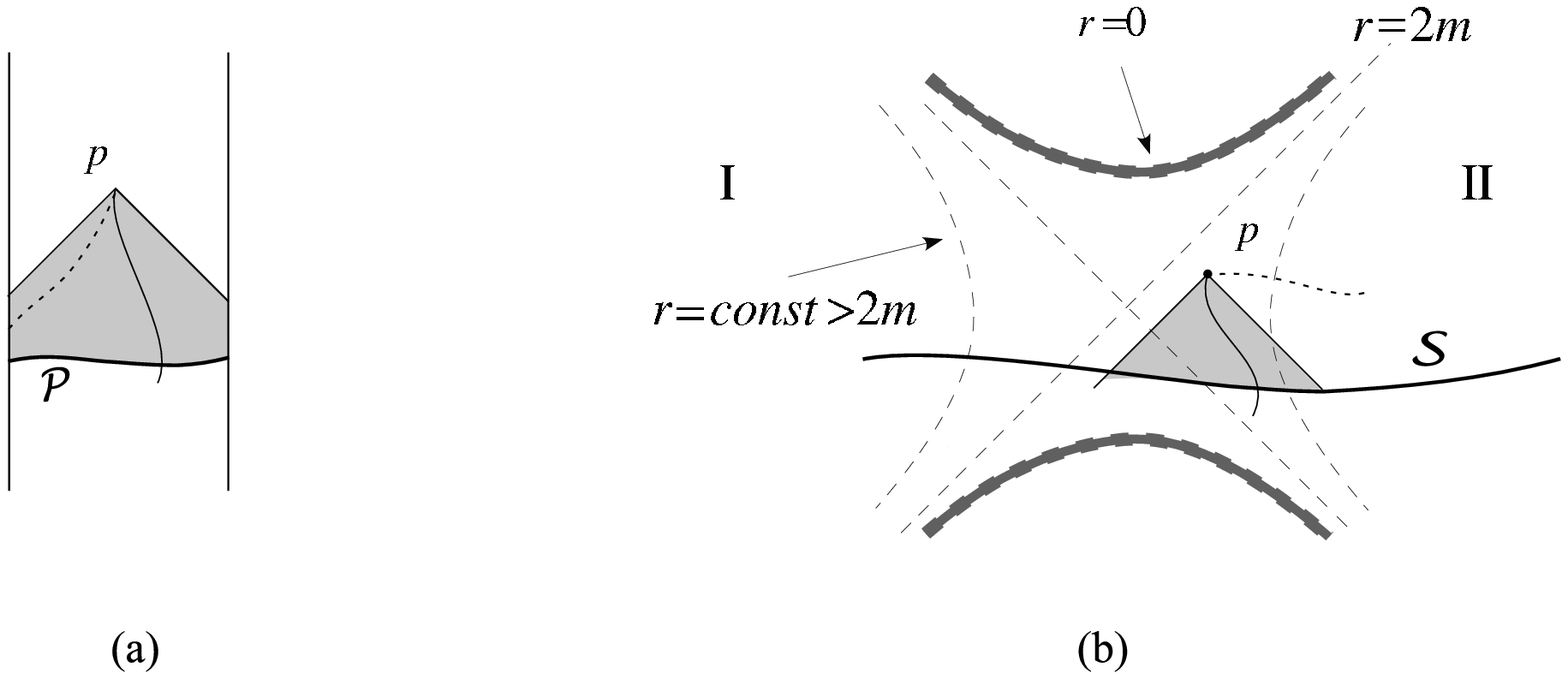}{\label{fig:3}\emph{a.} In the anti-de~Sitter space
some `unexpected' (i.~e.\ unknown at $\mathcal P$) information can
come to $p$ from infinity along the (timelike) dashed curve. \emph{b.}
In  the Schwarzschild spacetime everything, that reaches $p$, either
originates at $\mathcal S$, or moves with a superluminal speed.}

Let us call a system (i.~e.\ a set of fields, and/or
particles) \emph{predictable} if there is a spacelike surface such that
the  state of the system on that surface uniquely determines its state
at any future point. A theory is \emph{prognostic} if all systems
described by that theory are predictable.
A non-prognostic theory, i.~e.\ a theory unable to predict the outcome
of an experiment, is of doubtful practical value.

The importance of globally hyperbolic spacetimes lies in the fact that
all usual relativistic theories (mechanics, electrodynamics, etc.) are
prognostic in such spacetimes. If the
knowledge of everything that takes place at $\mathcal S$ is
insufficient for determining the state of the system at $p$, it means
that the necessary additional information either comes to $p$ just out
of nowhere (and so the corresponding theory is indeterministic), or
reaches $p$ after propagating along spacelike curves (only such curves
can avoid meeting $\mathcal S$). In the latter case the theory admits
superluminal signalling.

The reverse is also true: in a prognostic theory information does not
propagate faster than light. Indeed, take a point $p'$ slightly to the
future from $p$ and consider the set $I^-(p')$ of all points reachable
from $p'$ by past-directed timelike curves. It turns out that
$I^-(p')$ is globally hyperbolic, if so is the whole spacetime (this
is not obvious from Definition~\ref{def}, but can be easily
established from another definition of global
hyperbolicity~\cite{HawEl}, the equivalence of which to ours is a
highly non-trivial fact). Thus in a prognostic theory a state of the
system in $p$ is uniquely determined by what takes place in $I^-(p')$
(in fact, even by those events that lie on its Cauchy surface). In
this sense no information from $M-I^-(p')$ is available to an
observer in $p$. Our assertion then follows by continuity.

\parnum{Example.}  Consider a field obeying the equation
$$
 (\dalamb +\mu)\phi=0.
$$
This is a second-order linear hyperbolic equation. The Cauchy problem
for such equations is known to be well-posed in any globally
hyperbolic spacetime~\cite{HawEl}. So, $\phi$ is predictable and hence, in
particular, is unsuitable for superluminal signaling (at least on the
classical level). Which is noteworthy, because, when $\mu>0$, such a
field is traditionally referred to as `tachyonic'.

When a spacetime is globally hyperbolic and all material fields are
predictable\footnote{Actually, for a rigorous proof a few additional
mild assumptions are necessary, see chapter 7 in Ref.~\cite{HawEl}.},
general relativity itself is also a prognostic theory: the metric of a
spacetime is uniquely determined by the Einstein equations, if it is
known on a Cauchy surface.

\subsection{The general case}

If the condition of global hyperbolicity is relaxed the  situation
with predictability is \emph{much worse}. Let us illustrate it by
a few simple examples.

\kartiv{30pc}{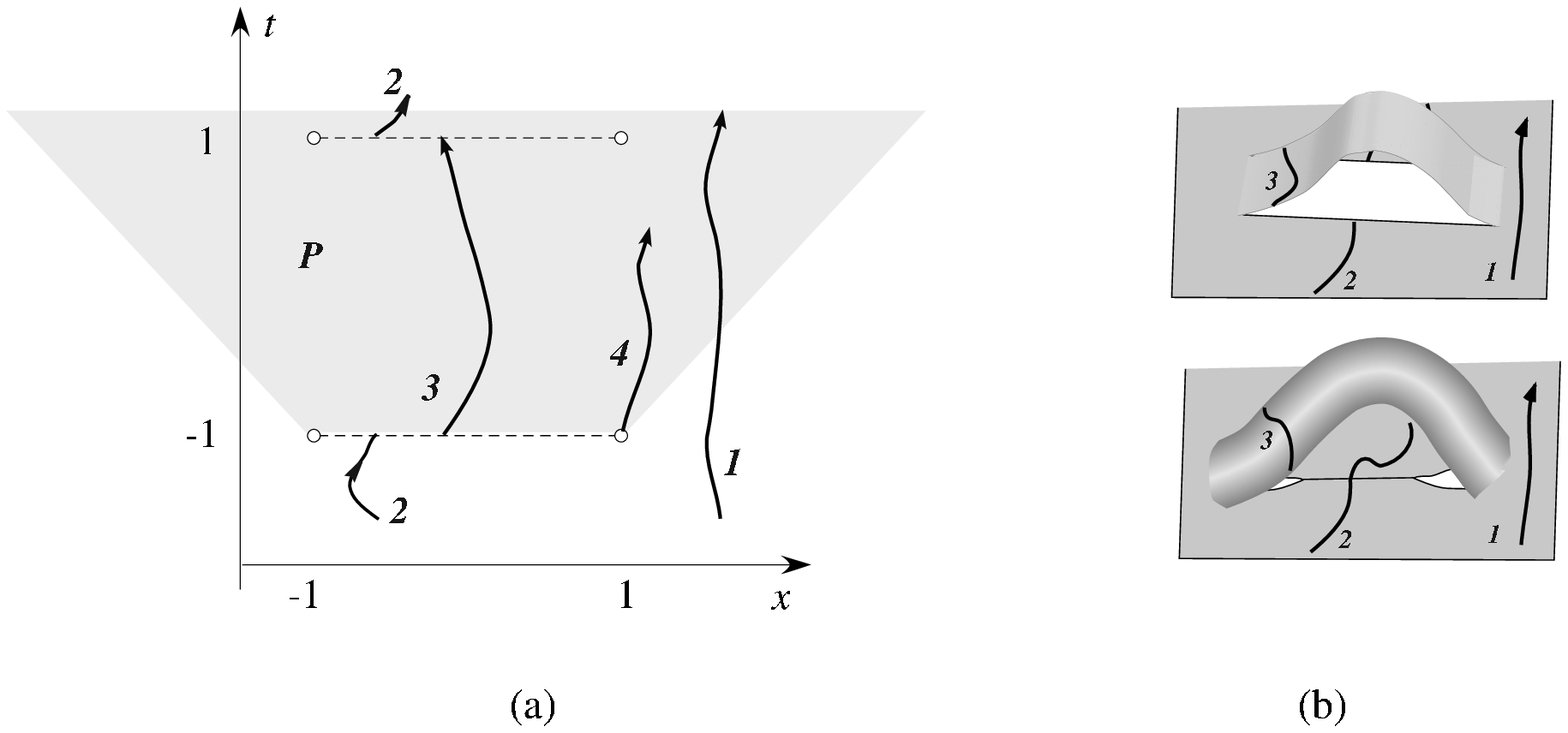}{The banks of the cuts (which are depicted by the
dashed lines) are glued so that curve 2 is actually continuous and
curve 3 --- closed.\label{fig:DP}}

Remove  the  points $(t=\pm1, \;x=\pm1)$ from the Minkowski plane.
Next, make the cuts along the two spacelike segments $\{t=\pm1,
\;-1< x <1\}$ connecting them (see Fig.~\ref{fig:DP}a).
 Now, preserving their orientation, glue the banks of the cuts --- the
upper bank of the lower cut to the lower bank of the upper cut and
vice versa.  The obtained spacetime $M$  --- called the
Deutsch--Politzer (DP) space --- is a combination of a cylinder and a
plane with two holes (see Fig.~\ref{fig:DP}b). The holes result from
excising the `corner' points (we cannot glue them back into the
spacetime) and form singularities of a somewhat unusual
kind\footnote{In many respects they resemble the conical singularity
from Example 1.}: the spacetime is flat and thus the singularities are
not associated with the divergence of curvature, or its derivatives. In
addition to singularities the spacetime contains closed causal curves,
which makes it one of the simplest models of the time
machine~\cite{Deu,Pol}. These are the images of the causal curves
which in the original (Minkowski) space start from the lower segment
and terminate at the upper.

\parnum{Remark.} We described the DP space as a result of some
`cut-and-paste' procedure. This, however, gives \emph{no} grounds at
all to consider it as something defective and unphysical. In fact, it
is exactly as `physical' as the Minkowski space, being the solution of
the Einstein equations with exactly the same (zero) source and exactly
the same initial conditions (see below). If desired, one could
start from the Deutsch--Politzer spacetime and obtain the Minkowski
space by a similar surgery.

The \emph{whole} spacetime $M$ is not globally hyperbolic (by
property~\ref{prop2}). However, all its `pathologies', i.~e.\
singularities and time loops, are confined to a region $P$ (in
Fig.~\ref{fig:DP}a it is shown gray) bounded by null geodesics
emerging from the lower `holes'. The remaining part $N\equiv M -
\overline{P}$ of $M$ is a `nice' spacetime. Specifically, $N$,
considered as a spacetime (see remark~\ref{rem:intr}),  \emph{is}
globally hyperbolic, the line $\mathcal S_N=\{p\colon\:t(p)=-2\}$
being one of its Cauchy surfaces.  We have thus a spacetime that
`looses' its global hyperbolicity in the course of evolution. This
lost, as can be easily seen, is fatal for the predictability of material
systems.
\parnum{Example.} Suppose our subject matter are pointlike sterile
particles, which is, probably, the simplest possible subject. Let us
fix the simplest possible initial data --- no particles at all at
$\mathcal S_N$. It is easy to check that neither these initial data,
nor the equations of motion\footnote{Nor, of course, any their
consequences, like the energy conservation.}  give us any clue about
what will happen later. Possibly the spacetime will always remain
empty. But exactly as well later there may appear some particles. Such
`new' particles (`lions' in terminology of
\cite{ttrpar}) may have the world lines emerging from a singularity (as
with particle 4), or rolled into a circle (particle~3).

Moreover,  it may happen that quite an innocent  Cauchy problem in a
non-globally hyperbolic spacetime does not have \emph{any solution}
whatsoever.  This fact is known as the `time
travel paradox'~\cite{ttrpar}.
\parnum{Example.}\label{expar} Consider a spacetime which is
constructed
exactly as the Deutsch--Politzer space, except that before gluing the
cuts, one of them is twisted by 180$^\circ$ (see
Fig.~\ref{fig:tDP}a).
\kartiv{30pc}{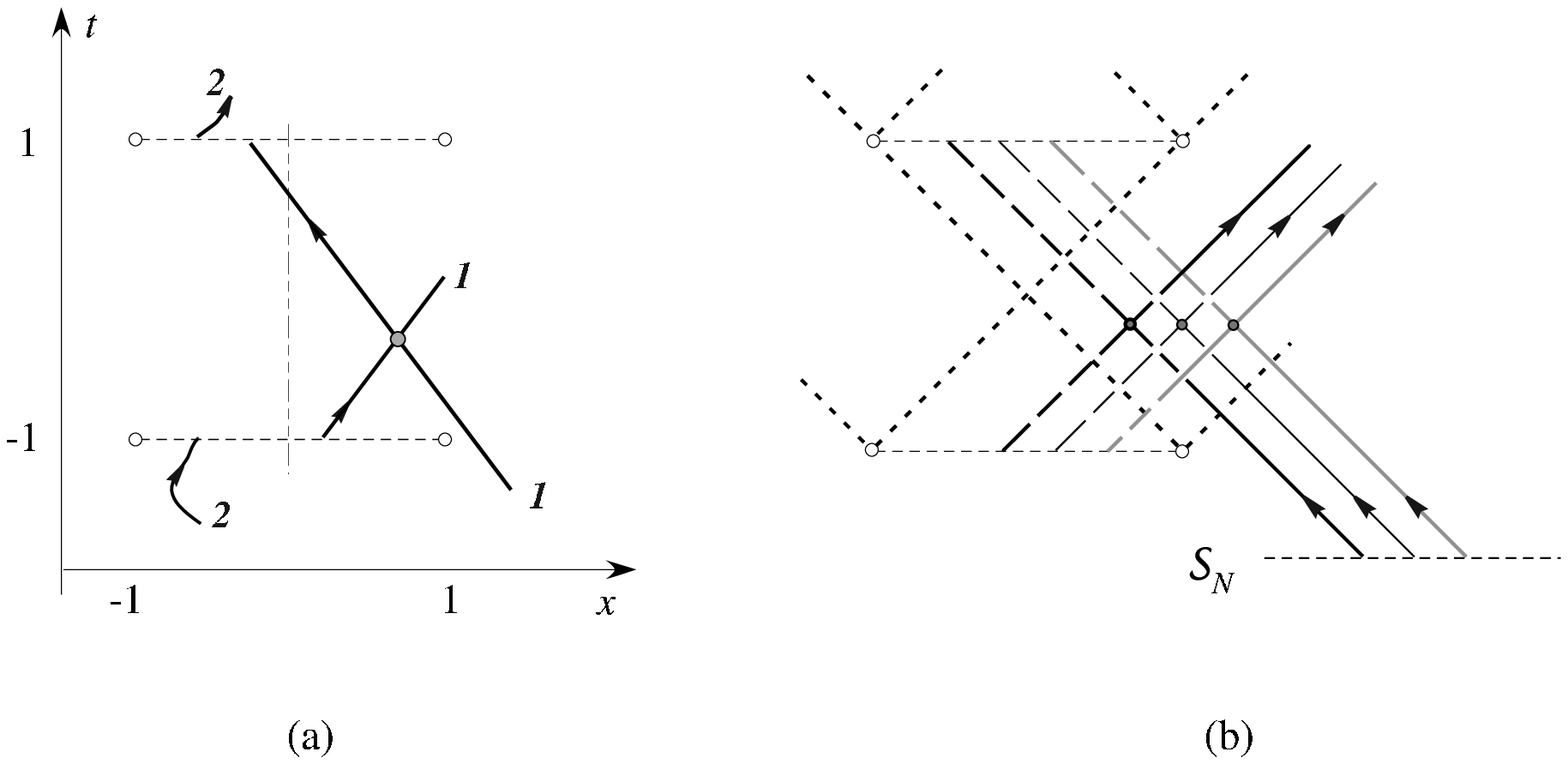}{\emph{a.} The twisted  Deutsch--Politzer space. The
banks of the horizontal cuts are glued so that both curves (1 and 2)
are smooth. \emph{b.} The time travel paradox. The trajectories of
all possible (given there are only 3 particles at $\mathcal S_N$)
particles are shown. The world lines of the `initial' particles cannot
be continued  in agreement with the laws of
interaction.\label{fig:tDP}}
Such a spacetime --- called the twisted
DP space --- is especially convenient for constructing paradoxes,
because the world lines even of free falling particles have
self-intersections. Our concern is with the behaviour of massless
particles  whose dynamics is determined by the following rules. Each
particle is characterized by two integer parameters --- we shall call
them `color' and `flavor' --- that take three and two values,
respectively. The particles are assumed to be sterile with a
\emph{single} exception: if two identical (i.~e.\ with the same values
of both parameters) particles meet, they both change their flavor.
Suppose now that at some moment $t<-1$ there are three particles of
different colors, all moving to the left, as shown in
Fig.~\ref{fig:tDP}b. If the spacetime remained globally hyperbolic at
later times, these  initial conditions would just determine the
evolution of the particles, but not in this case. In the twisted DP
space  \emph{no} evolution (governed by the formulated laws
of interaction) is consistent with the initial data~\cite{ttrpar}.

The situation with the evolution of the spacetime geometry is also bad in
the general case. Note, for example, that any sufficiently small
region of the DP space --- including the vicinity of $\mathcal S_N$
--- is isometric to the corresponding region of the Minkowski space.
Which means, in particular, that if the Minkowski space is a solution
of some Cauchy problem (i.~e. of a system \{some local --- Einstein's,
say --- equations +  conditions on $\mathcal S_N$\}), then the
DP space is also a solution of the same Cauchy problem. And this is
always the case: whenever a Cauchy problem in general relativity has a
globally hyperbolic solution $M$, it also has infinitely many
non-globally hyperbolic solutions (remove from $M$ a
two-sphere $\mathbb S$ lying to the future from $\mathcal S$; the
double covering of $M-\mathbb S$ is an example of such `spurious'
solution). Within classical relativity (complemented, if desired, with
any additional local laws) all these solutions are equipollent.

\section{Cosmic censorship hypothesis}

Summing up the preceding  we can state that physics is a
nice (prognostic) theory if it is known somehow that our spacetime
is globally hyperbolic. Otherwise, not so much can be said about
anything whatsoever. Strictly speaking, in the general case we cannot
give a substantiated answer even to the question: Where a car, moving
at 60$\,$km per hour, will find itself in 20 minutes? The honest
answer would be: If the car is lucky enough it will pass 20$\,$km, but
exactly as well it may happen that (after 10 minutes, say) a
singularity will appear out of thin air (as in the DP case) and the
car will  vanish in it, or will  be attacked by a monster that will
have emerged from the singularity, etc. All such possibilities are in
perfect agreement with both the initial data and the known physical
laws.

In everyday life people cope with this difficulty by assuming
(\emph{implicitly} as a rule) that the spacetime, indeed, is  globally
hyperbolic. So, it seems tempting to solve the whole problem by just
explicitly adopting global hyperbolicity as an additional
\emph{postulate}.  The spacetimes of the DP type would then be ruled
out and general relativity would consider only globally hyperbolic
maximal solutions of the Einstein equations as appropriate models of
the universe.

Such a program, however, immediately meets two problems. The first is
of the philosophical nature: it is hard to justify such a
\emph{non-local} postulate (indeed, as we saw in example~\ref{expar}, our
ability to perform some experiments \emph{now} may depend on whether
or not a causal loop will appear  somewhere in the
\emph{future}). The second problem is more serious: this new postulate
can come into contradiction with the `old' ones. Spacetimes are
conceivable (see below) where the lost of global hyperbolicity is an
\emph{inevitable} consequence of the Einstein equations (for specific
initial conditions, of course, and with maximality required). So, the
best one can hope is that such situations are impossible in
`realistic' circumstances. And it is this hope --- known as the
Cosmic censorship hypothesis --- that is endangered by the
wormholes.\footnote{If one discovered how to create a timelike
singularity, it also would violate the Cosmic censorship hypothesis.
This possibility, however, is much better known (see, for example,
in~\cite{Brady}) and I shall not discuss it here}
\section{Wormholes}
Pick  in  the Minkowski space two close  cylinders $C_1$ and $C_2$, of
which the second has a bend between, say, $t=-1$ and $t=1$. Except for
that bend both cylinders must be parallel to the $t$-axis (see
Fig.~\ref{fig:wh}a). On the boundaries of the cylinders $\mathcal
B_{1,2}\equiv\Bd C_{1,2}$ define a function $\tau$ as follows:
$\tau(p)$ is the length of the longest (recall that the metric is
Lorentzian) timelike curve that lies in $\mathcal B_{1,2}$ and
connects $p$ with the surface $t=-2$. Clearly, $\tau(p)=t(p)$ for all
$p\in\mathcal B_{1}$, but not for $p\in\mathcal B_{2}$. Now remove
the interiors of the cylinders and identify the boundaries $\mathcal
B_1$ and $\mathcal B_2$ so that
\kartiv{30pc}{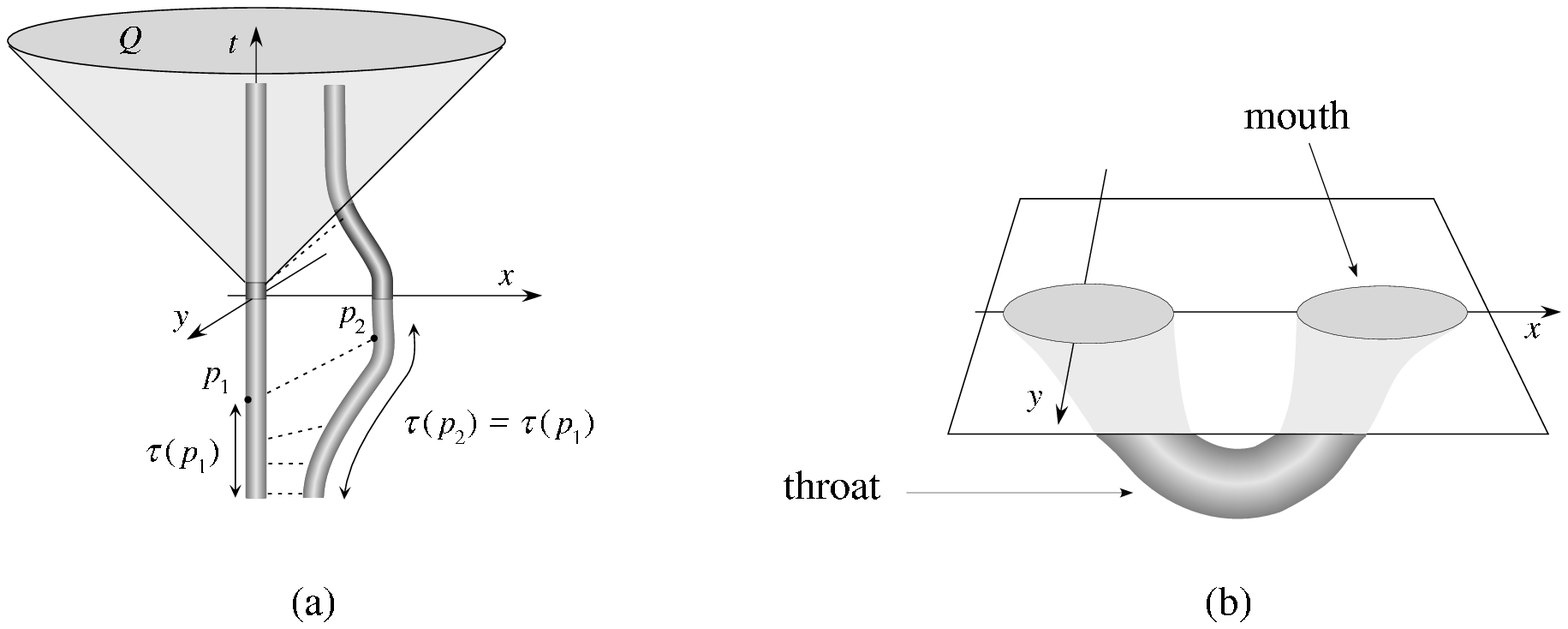}{When the identification is performed, the dashed
lines become circles. Note that as time goes these lines get more and
more tilted.\label{fig:wh} Finally, one of them (the one lying on the
boundary of the gray cone) becomes null and the first closed causal
curve appears.}
\begin{equation}\label{eq:rule}\tag{$*$}
p_1=p_2\quad\Rightarrow\quad \tau(p_1)=\tau(p_2),\qquad
p_{1,2}\in\mathcal B_{1,2}.
\end{equation}
Finally, smooth out the junction by curving appropriately a close
vicinity of $\mathcal B_1=\mathcal B_2$ (to remove the discontinuities
in derivatives of the metric).

To see what thus obtained spacetime $M_W$ presents, consider its
section $\mathcal S_N=\{p\colon\:t(p)=-2\}$. According to the
procedure described above, $\mathcal
S_N$ is obtained from the Euclidean
space $\mathbb E^3$ by removing two open balls, identifying the
boundaries of the holes, and smoothing out the junction. So, $\mathcal
S_N$ is a wormhole (in the two-dimensional case, see
Fig.~\ref{fig:wh}b, we would call it a handle). The former holes are
called \emph{mouths} and the `conduit' connecting them the
\emph{throat}. The form of the throat depends on just how we have smoothed
out the junction, but --- and this is important --- it can be made
(almost) constant (that is what the condition \eqref{eq:rule} was
imposed for). So, $M_W$ describes a wormhole one of whose mouths is
somehow pushed away from the other (without changing the length and
the form of the throat) and then returned back~\cite{MTY}.

This spacetime, proposed  by Morris, Thorne, and Yurtsever
(MTY)~\cite{MTY}, received much attention, because at one time it was
believed to describe the creation of a time machine. Indeed, the
identified points of $\mathcal B_1$ and $\mathcal B_2$ initially
(i.~e.\ at $t=-2$) have the same $t$-coordinate, but later, according
to \eqref{eq:rule}, each $p_{1}\in\mathcal B_{1}$ is identified with a
$p_{2}\in\mathcal B_{2}$ such that $t(p_2)>t(p_1)$. As soon as
$t(p_2)-t(p_1)$ becomes greater than $x(p_2)-x(p_1)$  the identified
points turn out to be causally connected in the `initial' (i.~e.
Minkowski) space. Which means that $M_W$ contains closed causal
curves.

As with the DP space, the causal loops are confined to a region $Q$
(the gray cone in Fig.~\ref{fig:wh}a) that lies to the future of a
globally hyperbolic (when considered in itself) spacetime $L\equiv M_W
-\overline Q$. So, we again have a spacetime loosing its global
hyperbolicity in the course of evolution. However, there is also a
striking difference with the previous case. The additional postulate
--- to the effect that the whole (i.~e.\ maximal) spacetime must be
globally hyperbolic ---  only required that $N$ would evolve
into the Minkowski (and not  Deutsch--Politzer, say) space,
something that intuitively is quite acceptable and even appealing. But
$L$ \emph{does not} have globally hyperbolic extensions and therefore
makes us choose between maximality and global hyperbolicity.

Thus the wormholes (if they exist, if they are stable, if their mouths
can be separately moved, etc.) refute the Cosmic censorship
hypothesis. Just take a wormhole, push one of its mouths, and pull it
back. Whatever results\footnote{As has been proved recently, it need
not be a time machine: among the possible extensions of $L$ there
always are causal ones~\cite{eotm}.}, it will be in any case
non-globally hyperbolic.

\section{Global hyperbolicity protection}
How serious is  the wormhole hazard and do wormholes exist in nature,
in the first place?  The wormholes are often considered as
`a marginal idea'~\cite{Davies}, as something
`too exotic', that is, essentially, as something
that unlikely exists. The reasons for such a belief are not always
clear, but, at any rate, they are neither experimental, nor
theoretical.

The idea at the heart of general relativity is that the spacetime we
live in is a curved four-dimensional manifold. Which immediately
provokes a question: What is it so special in $\textrm{I\!R}^4$, that
one would believe it to be the only possible topology of the universe?
The answer (as of today, at least) is obvious --- nothing. One hardly
would be surprised, for example, if it turn out that we live in a
spatially closed universe.

The argument: ``We haven't ever seen any wormholes, so they don't
exist" does not, of course, stand up. The wormholes do not shine like
stars and are not supposed to be `seen'. The presence of a wormhole
surely would strongly affect the surrounding matter, but to what
observable effects it must lead, is yet to be learned. Some progress
in this direction has been achieved by  Cramer
\emph{et~al.}~\cite{lens}, who noticed that the gravitational lensing
of wormholes may differ from that of stars. By Birkhoff's theorem the
gravitational field of a static spherically symmetric wormhole (in the
empty region around it) is that of a pointlike massive source. The
value $m$ of the corresponding mass depends on what wormhole is
considered and at present there are no reasons to regard any $m$ as `more
realistic'. It is important, however, that
\emph{in particular} $m$ may be negative. Such wormholes would act more
like diverging lenses in contrast to the stars with their positive
masses. These considerations enabled  Torres
\emph{et~al.}~\cite{bursts} to find some (though not too restrictive)
bounds on the possible abundance of wormholes with negative $m$.

In a globally hyperbolic spacetime only primordial wormholes may exist
(see property~\ref{prop1}). So, the most direct way to support the Cosmic
censorship hypothesis would be to find a mechanism excluding their
existence at the onset of classical physics (i.~e.\ at the end of the
Planck era). Needless to say that no such mechanism is known yet.

Another way out would be to prove that  realistic wormholes cannot be
\emph{traversable}~\cite{Tho}, i.~e.\ large and long-lived enough to pass a
macroscopic object through the throat\footnote{A standard example of a
\emph{non}-traversable wormhole is the Schwarzschild solution. Even
though there are two asymptotically flat regions (I and II in
Fig.~\ref{fig:3}b) connected with a `bridge', a traveler cannot get
from one of them to the other because the bridge collapses too fast.}.
Indeed, the Einstein equations ensure~\cite{FSW} that to be
traversable a wormhole must be maintained by the `exotic matter' (i.~e.\
the matter whose  energy density is negative in some points). Which
means, in fact, that at the classical level the traversable wormholes
\emph{are} prohibited (except for a few rather exotic possibilities
such as the classical scalar field~\cite{V}, or ghost
radiation~\cite{hay} as a source). So, if it turns out that quantum
effects (which are known to produce negative energy densities as, for
example, in the Casimir effect~\cite{BirDav})  also fail to support
macroscopic wormholes, the problem would be solved. Actually, however,
the quantum fields seem to be well suited for the task. In particular,
a wormhole was found such that the zero point fluctuations of the
electro-magnetic or neutrino fields in its throat produce just enough
exotic matter to sustain the wormhole~\cite{MWH}.

Lastly, it may happen that traversable wormholes do exist, but,
nevertheless, the MTY scenario does not work because the mouths cannot
be moved appropriately. The point is that in the spacetimes of that type
there always exist `almost closed' null geodesics. In
Fig.~\ref{fig:wh}a such a geodesic goes from the left mouth,  enters
the right one, comes again from the left, etc. Though always
remaining in $L$, a photon with this world line makes infinitely many
trips between the mouths getting more and more blue~\cite{Cl}. This
may indicate that the process is unstable in the sense that an
occasional photon can prevent one from bringing the mouths close
enough --- the closer are the mouths the stronger is the resistance
offered by the photon. Of course, classically such an instability can
be cancelled by just placing an opaque screen between the mouths, but
one does not expect the
\emph{quantum} modes to be counteracted as easily. Which suggests that
the MTY process may suffer quantum instability. To verify this
hypothesis it is instructive to study the behavior of the vacuum
expectation of the stress-energy tensor near $\Bd L$ and to check
whether or not it is bounded~\cite{Frolov}. This has been done in a
few simplest (two-dimensional) cases and no evidence for the quantum
instability was found --- exactly as with the Minkowski space the
energy density in some cases blows up~\cite{Yurt1} and in other
cases~\cite{Noqins,Sush} does not.

\end{document}